\documentclass[prc,aps,twocolumn]{revtex4}
\usepackage{graphicx}
\usepackage{color}
\begin{document}
\title{An overview of symmetric nuclear matter properties from chiral interactions up     
to fourth order of the chiral expansion} 

\author{Francesca Sammarruca and Randy Millerson}
\affiliation{Department of Physics, University of Idaho, Moscow, ID 83844, USA}

\begin{abstract}
We present and discuss predictions for a cross section of bulk and single-particle properties in symmetric nuclear matter
 based on recent high-quality nucleon-nucleon potentials at N$^3$LO and including all subleading three-nucleon forces.
We begin with the equation of state and its saturation properties and proceed to the single-nucleon potential.  We also explore short-range correlations as seen through the defect function.
The various predictions which we present  have a  common  foundation in an internally consistent ab initio approach.
\end{abstract}
\maketitle 
        
\section{Introduction} 
\label{Intro}

Constructing the equation of state (EoS) of infinite nuclear matter microscopically from state-of-the-art few-body interactions remains an
important theoretical challenge in nuclear physics. The EoS gives fundamental
insight into effective nuclear forces in the medium, and is a crucial input in a variety of fields, ranging from heavy-ion (HI) reactions to astrophysical processes.

High-precision meson-theoretic or phenomenological interactions~\cite{Mac01,Sto94,WSS95} are still frequently employed in contemporary calculations of nuclear matter, structure, and reactions. However, in those models of the past,
three-nucleon forces (3NFs), or more generally $A$-nucleon forces with $A>2$, have only a loose 
connection with the associated two-nucleon force (2NF)~\cite{Mac17}. Furthermore, there exists no clear 
scheme to quantify and control the theoretical uncertainties. Chiral effective field theory (EFT)~\cite{ME11,EHM09,MS16}, on the other hand, provides a systematic approach for constructing
nuclear many-body forces, which emerge on an equal footing~\cite{Wei92} with two-body forces, and 
for assessing theoretical uncertainties through an expansion controlled by the
``power counting"~\cite{Wei90} method. Furthermore, chiral EFT maintains consistency with the symmetries and of the underlying fundamental theory of strong interactions, quantum 
chromodynamics (QCD), and the breaking of those symmetries.

For the reasons described above, chiral EFT has evolved into the authoritative approach for 
developing nuclear forces, and modern applications have focused on few-nucleon 
reactions~\cite{Epe02,NRQ10,Viv13,Gol14,Kal12,Nav16}, the structure of light- and medium-mass nuclei~\cite{Coraggio07,Coraggio10,Coraggio12,Hag12a,Hag12b,BNV13,Gez13,Her13,Hag14a,Som14,Heb15,Hag16,Car15,Her16,Hol17,Sim17,Mor17},
infinite matter at zero temperature~\cite{HS10,Heb11,Baa13,Hag14b,Cor13,Cor14,Sam15,Dri16,Tew16,MS16,Hol17}
and finite temperature~\cite{Wel14,Wel15}, and nuclear dynamics and response functions \cite{Bac09,Bar14,Rap15,Bur16,Hol16,Bir17,Rot17}.
Although satisfactory predictions have been obtained in many cases, specific problems persist.
These include the description of bulk properties of medium-mass nuclei, which typically
exhibit charge radii that are too small~\cite{Lap16} and binding 
energies that are highly sensitive to the choice of nuclear force and often turn out to be 
too large~\cite{Bin14}. 
More recently, it has been observed that chiral two- and three-nucleon interactions (at N$^2$LO and at N$^3$LO) which have been found to predict realistic binding energies and radii for a wide range of finite nuclei (from p-shell nuclei up to nickel isotopes) are unable to saturate infinite nuclear matter~\cite{Huether+2020}. On the other hand, it has been shown that, when the fits of the $c_D$ and $c_E$ couplings of the chiral three-nucleon interactions include the constraint of nuclear matter saturation in addition to, as is typically the case, the triton binding energy, medium-mass nuclei are underbound and their radii are sytematically too large~\cite{Hoppe19}.

This has led some groups to fit the low-energy constants that parametrize
unresolved short-distance physics in chiral nuclear forces directly to the properties of 
medium-mass nuclei~\cite{Eks15} and, indeed, better predictions for other isotopes are then 
obtained. However, in the {\it ab initio} spirit, one would prefer a genuine microscopic approach in which the 2NF is fixed 
by two-nucleon data and the 3NF by three-nucleon data, with no further fine tuning. Applications to 
systems with $A>3$ would then be true predictions, though possibly with large uncertainties.

In Ref.~\cite{EMN}, high-quality soft chiral $NN$ potentials from leading order to 
fifth order in the chiral expansion were constructed. These interactions are more consistent than those 
constructed earlier~\cite{EM03,chinn5,ME11}, in the sense that the same power counting scheme and 
cutoff procedures are used at all orders. For these potentials, the very accurate $\pi N$ low-energy 
constants (LECs) determined in the Roy-Steiner analysis of Ref.~\cite{Hofe+} are applied. The 
uncertainties associated with these LECs are so small that variations within the errors have negligible 
impact on the construction of the potentials. These potentials are soft and have good perturbative behavior, as
 demonstrated in the investigations of Refs.~\cite{Hop17,DHS19}.

In a recent work~\cite{SM21}, we concentrated on the neutron matter (NM) EoS and the density dependence of the symmetry energy with chiral 2NFs and 3NFs up to N$^3$LO,  order-by-order and with proper chiral uncertainty quantification. Our main focal point was the symmetry energy, which we discussed in relation to recent empirical constraints~\cite{Tsang21}. In the present work, we wish to address several aspects related to the EoS of symmetric nuclear matter (SNM), from bulk to single-particle properties. First, we will show order-by-order predictions for the EoS and quantify the truncation error.
 In this way, we will be able to assess the level of agreement with previous work based on the same 2NF~\cite{DHS19}, where a different many-body method is utilized.

Having addressed bulk properties, we will study the impact of 3NFs on the single-particle potential. Single-particle energies, often parametrized in terms of effective masses, provide insight into both density and momentum dependence of the in-medium interaction, and  are an important part of the  input for transport calculations. 

Single-particle properties are impacted by short-range correlations (SRC), which we will address next. We will explore SRC in nuclear matter as seen through the correlated {\it vs.} the uncorrelated wave functions. In particular, we will investigate the impact of complete 3NFs at N$^3$LO on central and tensor correlations.
Short-range correlations have been at the forefront of recent discussions.
Claims that momentum distributions in nuclei, with particularly emphasis on SRC, can be measured have stimulated considerable interest in the subject. These are not new discussions, but they have recently resurfaced in conjunction with inclusive or exclusive high-momentum transfer electron scattering experiments~\cite{CLAS,CLAS2,CLAS3,Pia+,Eg+06,Shneor,Subedi,Baghda,Pia13,Korover,Hen+17,CT+,Atkwim19}.
We will include  a brief discussion of the issue.

The manuscript is organized as follows: in Sec.~\ref{ff} we briefly
summarize the main features of the 2NFs and 3NFs employed in this work.
The reader is referred to Ref.~\cite{EMN} for a complete and
detailed description of the 2NF. In Sec.~\ref{IV} we present and discuss
 a variety SNM properties. 
Our conclusions are summarized in Sec.~\ref{Concl}, along with near-future plans.
     
\section{Few-nucleon forces}  
\label{ff} 

\subsection{The two-nucleon force}
\label{II}

The $NN$ potentials employed in this work are part of a set that spans five orders in the chiral EFT expansion, 
from leading order (LO) to fifth order (N$^4$LO), with the same power counting scheme and 
regularization procedures applied through all orders.
Another novel and important aspect in the construction of these improved potentials is the fact 
that the long-range part of the interaction is fixed by the $\pi N$ LECs as determined in the very accurate analysis of Ref.~\cite{Hofe+} -- in practice,
errors in the $\pi N$ LECs are no longer an issue with regard to uncertainty quantification.
Furthemore, at the fifth (and highest) order, the $NN$ data below pion production threshold 
are reproduced with high precision ($\chi ^2$/datum = 1.15).

Iteration of the potential in the Lippmann-Schwinger equation, and the fact that we are building a low-momentum expansion, require                        
cutting off high-momentum components.      
This is accomplished through the application of a regulator function for which we choose the non-local form                                
\begin{equation}
f(p',p) = \exp[-(p'/\Lambda)^{2n} - (p/\Lambda)^{2n}]     \; .
\label{reg}
\end{equation}
For the reasons mentioned in Sec.~\ref{Intro}, we will employ the softer version of these potentials, with cutoff $\Lambda$ = 450 MeV.

\begin{figure*}[!t] 
\centering
\hspace*{-1cm}
\includegraphics[width=9.0cm]{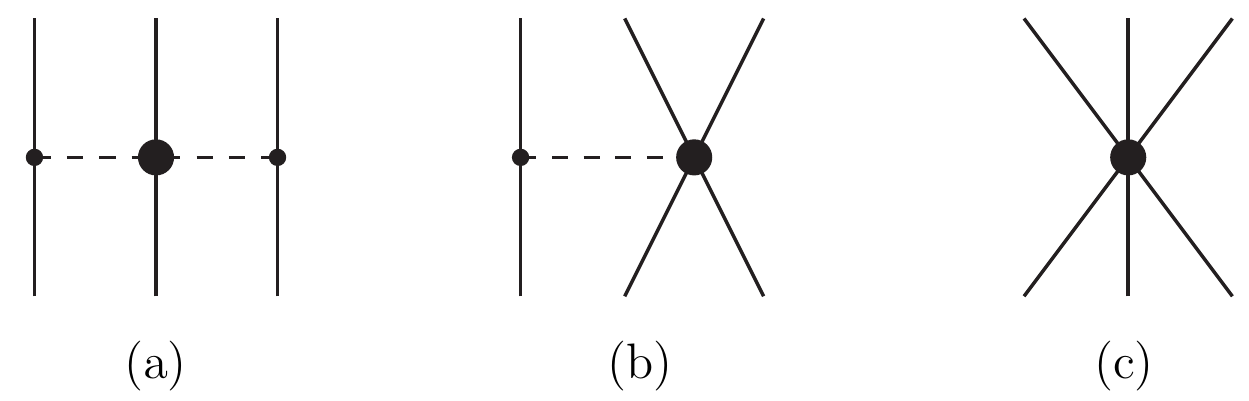}\hspace{0.01in} 
\vspace*{-0.05cm}
 \caption{ Diagrams of the leading 3NF: (a) the long-range 2PE, depending on the LECs $c_{1,3,4}$; (b) the medium-range 1PE, depending on the LEC $c_{D}$; (c) the short-range contact, depending on the LEC $c_E$.}
\label{3nf_n2lo}
\end{figure*}

\subsection{The three-nucleon force} 
\label{III} 

\begin{figure*}[!t] 
\centering
\hspace*{-1cm}
\includegraphics[width=11.5cm]{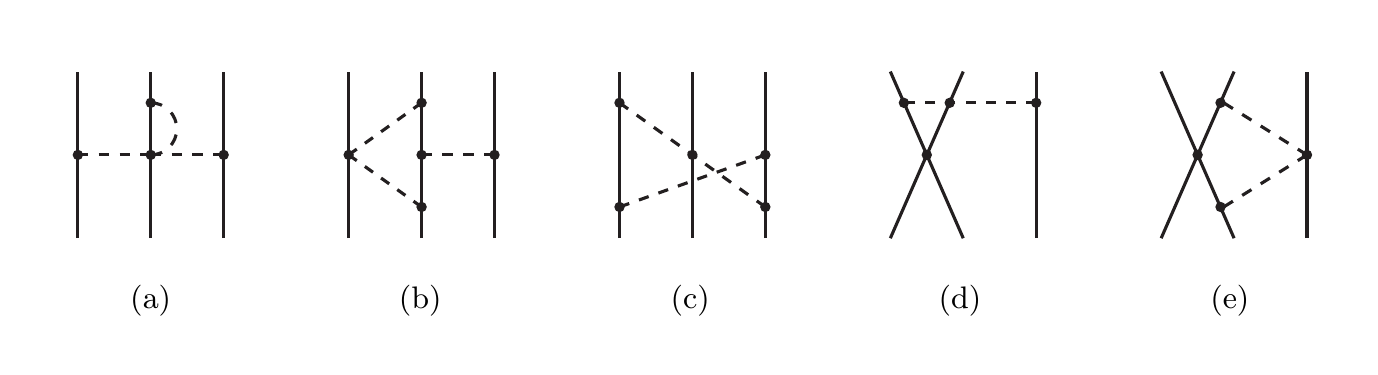}\hspace{0.01in} 
\vspace*{-0.5cm}
 \caption{ Some diagrams of the subleading 3NF, each being representative of a particular topology: (a) 2PE; (b) 2P1PE; (c) ring; (d) 1P-contact; (e): 2P-contact. Note that the 1P-contact topology makes a vanishing contribution.
}
\label{3nf_n3lo}
\end{figure*}

Three-nucleon forces first appear at N$^2$LO of the $\Delta$-less theory, which we apply in this work. At this order, the 3NF consists of three contributions~\cite{Epe02}: the long-range two-pion-exchange (2PE) graph, the medium-range one-pion-exchange (1PE) diagram, and a short-range contact term. We show the  topologies in Fig.~\ref{3nf_n2lo}.
In infinite matter, these 3NFs can be expressed in the form of density-dependent effective two-nucleon interactions as derived in Refs.~\cite{holt09,holt10}. They are represented in  terms of the well-known non-relativistic two-body nuclear force operators and, therefore, can be conveniently incorporated in the usual $NN$ partial wave formalism and the particle-particle ladder approximation for computing the EoS. The effective density-dependent two-nucleon interactions at N$^2$LO consist of six one-loop topologies. Three of them are generated from the 2PE graph of the chiral 3NF and depend on the LECs $c_{1,3,4}$, which are already present in the 2PE part of the $NN$ interaction. Two one-loop diagrams are generated from the 1PE diagram, and depend on the low-energy constant $c_D$. Finally, there is the one-loop diagram that involves the 3NF contact diagram, with LEC $c_E$.

 The 3NF at N$^3$LO has been derived~\cite{Ber08,Ber11} and applied in some nuclear many-body systems~\cite{Tew13,Dri16,DHS19,Heb15a}. The long-range part of 
 the subleading chiral 3NF consists of (cf. Fig.~\ref{3nf_n3lo}): the 2PE topology, which is the longest-range component of the subleading 3NF, the two-pion-one-pion exchange (2P1PE) topology, and the ring topology, generated by a circulating pion which is absorbed and reemitted from each of the three nucleons. 
The in-medium $NN$ potentials corresponding to these long-range subleading 3NFs in SNM are given in Ref.~\cite{Kais19}. The short-range subleading 3NF consists of (cf. Fig.~\ref{3nf_n3lo}): the one-pion-exchange-contact topology (1P-contact), which gives no net contribution, the two-pion-exchange-contact topology (2P-contact), and relativistic corrections, which depend on the $C_S$ and the $C_T$ LECs of the 2NF and are proportional to $1/M$, where $M$ is the nucleon mass. 
The in-medium $NN$ potentials corresponding to the short-range subleading 3NFs in SNM can be found in Ref.~\cite{Kais18}.

The LECs we use in this work, displayed in Table~\ref{tab1}, are from Ref.~\cite{DHS19}.  A technical remark is in place: when the subleading 3NFs are included, the $c_1$ and $c_3$ LECs are replaced by -1.20 GeV$^{-1}$ and -4.43 GeV$^{-1}$, respectively. This is because most of the subleading two-pion-exchange 3NF has the same mathematical structure as the leading one~\cite{KGE12} and thus, in practice, a large part of the subleading two-pion-exchange 3NF can be accounted for 
with a shift of the LECs equal to -0.13 GeV$^{-1}$ (for $c_1$), 0.89 GeV$^{-1}$ (for $c_3$), and -0.89 GeV$^{-1}$ (for $c_4$)~\cite{Ber08}.

\begin{table*}[t]
\caption{
Values of the LECs $c_{1,3,4}$, $c_D$, and $c_E$ 
for different orders in the chiral EFT expansion. The momentum-space
cutoff $\Lambda$ is equal to 450 MeV.
The LECs $c_{1,3,4}$ are given in units of GeV$^{-1}$, while
$c_D$ and $c_E$ are dimensionless.}
\label{tab1}
\begin{tabular*}{\textwidth}{@{\extracolsep{\fill}}ccccccccc}
\hline
\hline
  & $\Lambda$ (MeV) & $c_1$ & $c_3$ & $c_4$ &  $c_D$ & $c_E$  & $C_S$  & $C_T$ \\
\hline    
\hline
N$^2$LO & 450 &  --0.74 & --3.61 & 2.44  &  (a) 2.25 &   0.07 & -0.013000 & -0.000283 \\
     &     &              &      &                              & (b)  2.50 &   0.1 & &  \\
     &     &              &      &                              & (c)   2.75 &   0.13 & &  \\    
\hline 
N$^3$LO & 450 & --1.07 & --5.32 & 3.56  &  (a)  0.00 &   -1.32  & -0.011828 &  -0.000010 \\
      &     &              &      &                              &(b)   0.25  &   -1.28 & &  \\
     &     &              &      &                              &  (c)  0.50 &   -1.25 &  & \\
\hline
\hline
\end{tabular*}
\end{table*}

\section{Symmetric nuclear matter} 
\label{IV} 

We perform microscopic calculations of nuclear matter with the interactions described above.
We compute the EoS using the nonperturbative particle-particle
ladder approximation, which generates the leading-order contributions 
in the traditional hole-line expansion. We compute 
the single-particle spectrum for the intermediate-state energies self-consistently,  keeping the real part.

\subsection{Order by order predictions for the EoS} 
\label{IVa} 

\begin{figure*}[!t] 
\centering
\hspace*{-1cm}
\includegraphics[width=7.5cm]{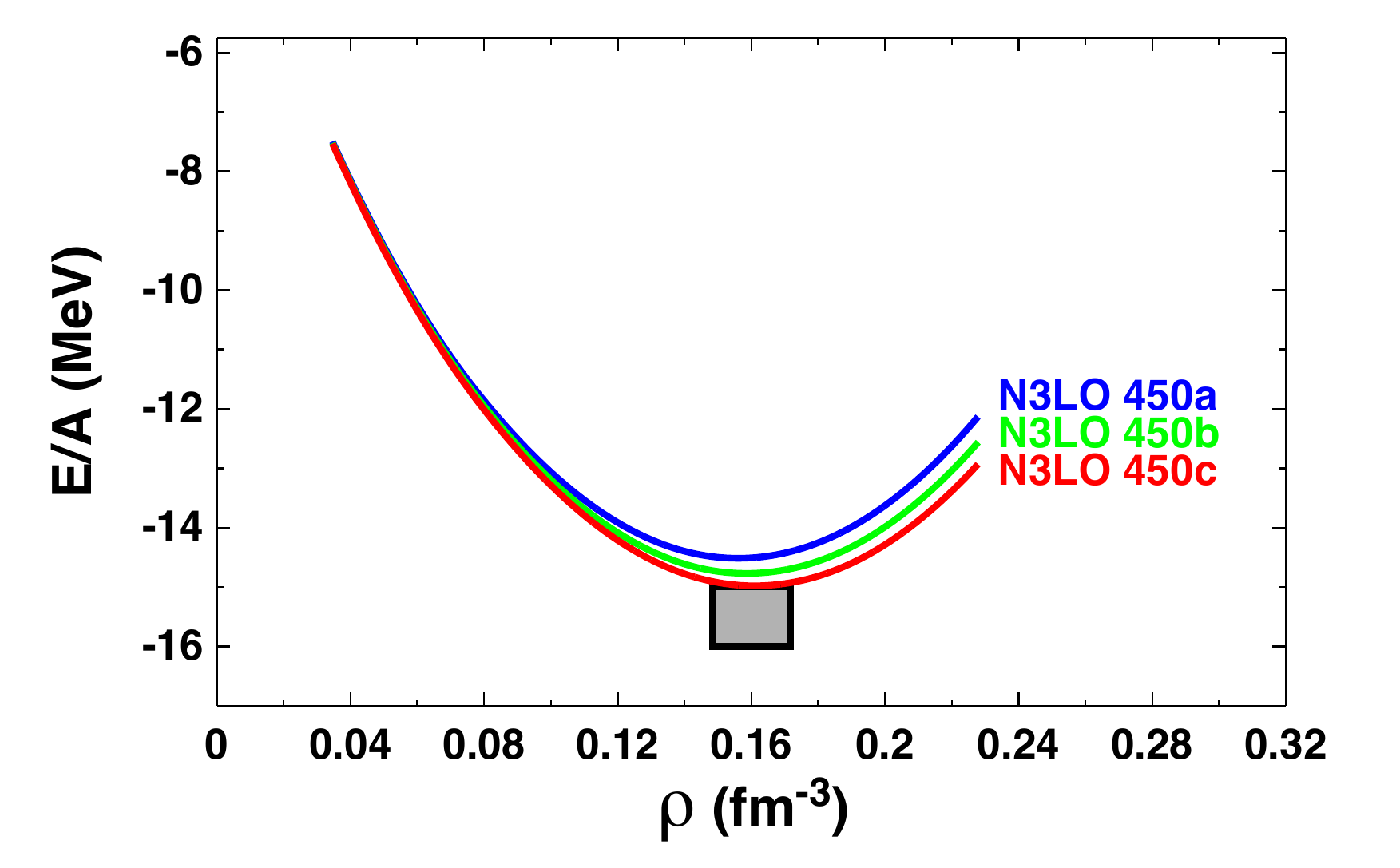}\hspace{0.01in} 
\vspace*{0.05cm}
 \caption{ Energy per particle as a function of density at N$^3$LO and cutoff equal to 450 MeV. The labels a, b, and c refer to the different sets of $c_D, c_E$ values given in Table~\ref{tab1}.
}
\label{eos_n3lo}
\end{figure*}

We begin with the study displayed in Fig.~\ref{eos_n3lo}. The curves are obtained with $\Lambda$ = 450 MeV  and the different sets of $c_D, c_E$ LECs displayed in Table~\ref{tab1}, of which set (c) produces the best saturating behavior.
  There is growing consensus that softer potentials are more likely to give favorable predictions in nuclear structure, although the problem of constructing an accurate $NN$ interaction that is successful in the intermediate-mass region  remains unsolved~\cite{NEM21}.

\begin{figure*}[!t] 
\centering
\hspace*{-1cm}
\includegraphics[width=7.5cm]{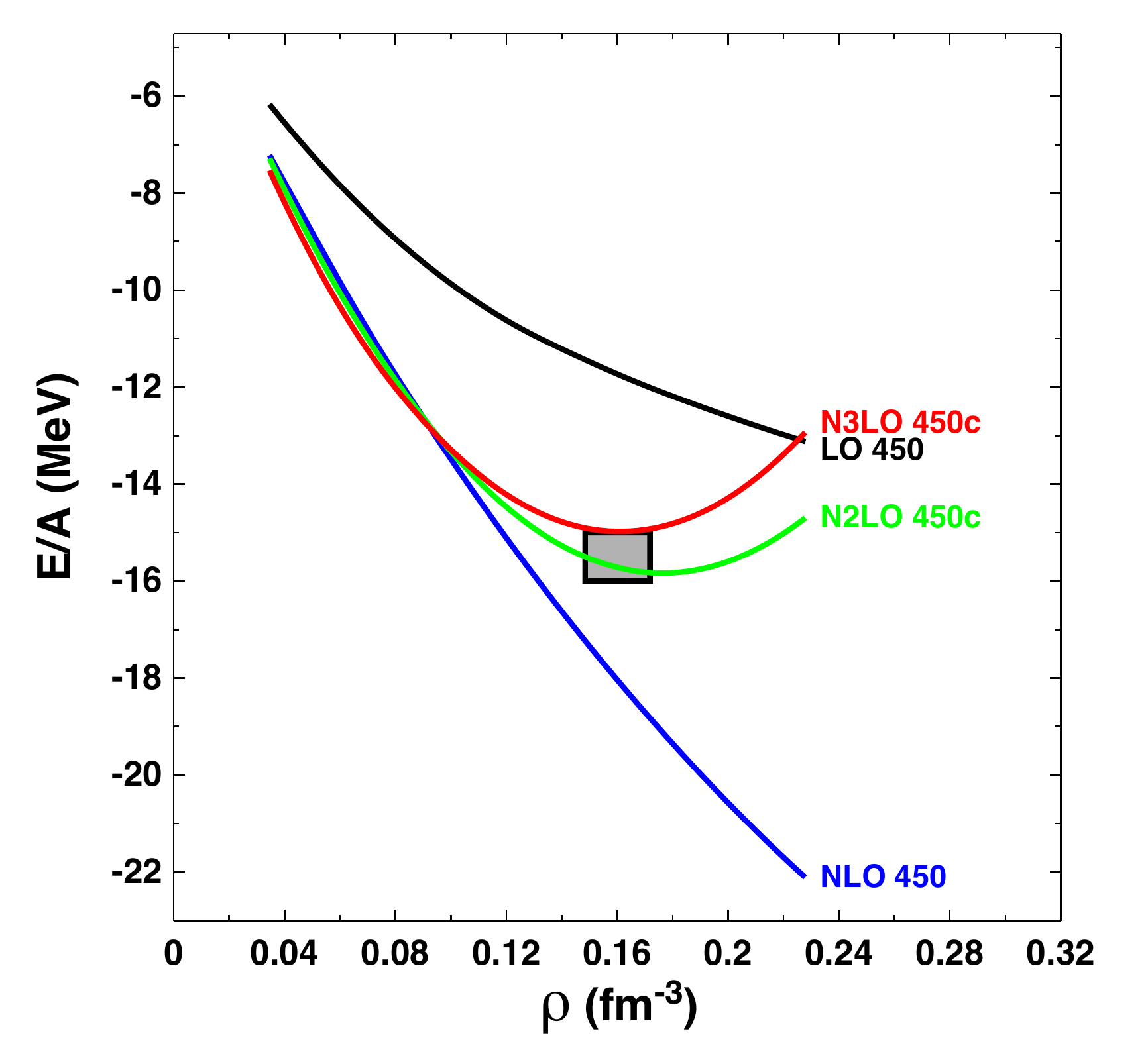}\hspace{0.01in} 
\vspace*{0.05cm}
 \caption{ Energy per particle as a function of density from leading to fourth order of chiral perturbation theory. The cutoff is fixed at 450 MeV.
}
\label{obo_c}
\end{figure*}

In Fig.~\ref{obo_c}, we show the energy per particle from leading to fourth order.  While  the EoS has already a realistic behavior at the first order where 3NFs appear (N$^2$LO), there is a definite improvement when moving to N$^3$LO, for both saturation density and energy. This is an important validation of the predictive power of the chiral EFT -- of course, $NN$ data and the three-nucleon system must be described accurately for any subsequent many-body predictions to be meaningful. 

Next, we discuss chiral uncertainties.
As pointed out in Sec.~\ref{II}, errors in the $\pi N$ LECs are no longer an issue with regard to uncertainty quantification. On the other hand, crucial to chiral EFT 
is the truncation error. If observable $X$ is known at order $n$ and at order $n+1$, a reasonable estimate of the truncation error at order $n$ can be expressed as the difference between the value at order $n$ and the one at the next order:
\begin{equation}
\Delta X_n = |X_{n+1} - X_n| \; ,
\label{del} 
\end{equation} 
since this is a measure for what has been neglected at order $n$.
To estimate the uncertainty at the highest order that we consider, we follow the prescription of Ref.~\cite{Epel15}. For an observable $X$ that depends on the typical momentum of the system under consideration, $p$, one defines $Q$ as the largest between $\frac{p}{\Lambda_b}$ and $\frac{m_{\pi}}{\Lambda_b}$, where $\Lambda_b$ is the breakdown scale of the chiral EFT, for which we assume 600 MeV~\cite{Epel15}. The uncertainty of the value of $X$ at N$^3$LO is then given by:
\begin{displaymath}
\Delta X = \max \{Q^5|X_{LO}|, Q^3|X_{LO} - X_{NLO}|,Q^2|X_{NLO} - X_{N^2LO}|, 
\end{displaymath}
\begin{equation}
Q|X_{N^2LO} - X_{N^3LO}| \} \; ,
\label{err}
\end{equation} 
where $p$ could be identified with the Fermi momentum at the density under consideration. To evaluate the truncation error for saturation parameters using Eq.~(\ref{err}), one might define a nominal ``saturation" density, say $\rho_0$ = 0.16 fm$^{-3}$, for all orders. On the other hand,  the EoS at LO and NLO do not exhibit a saturating behavior, thus,
 it may be more meaningful to consider the actual saturation densities for the EoS which do saturate (namely, those including 3NFs), especially for the purpose of evaluating the  incompressibility, which measures the curvature of the EoS at the minimum. Estimating (pessimistically) the truncation error at N$^3$LO as $|X_{N^3LO} - X_{N^2LO}| $, we find, for the saturation density at N$^3$LO, $\rho_0$ = (0.161 $\pm$ 0.015) fm$^{-3}$. 
Proceeding in the same way for the saturation energy and the incompressibility, we find, at N$^3$LO, $e(\rho_0)$ = (-14.98 $\pm$ 0.85) MeV, and $K_0$ = (216 $\pm$ 33) MeV.
Adopting, instead, the prescription  $|X_{N^3LO} - X_{N^2LO}|  \frac{Q}{\Lambda}$, where $Q$ is identified with the Fermi momentum at saturation density, the errors would be reduced by about 44\%.

Lastly, we note that our N$^3$LO(450) results for the energy per particle at saturation agree with those from Ref.~\cite{DHS19} within uncertainties.

\subsection{The single-particle potential} 
\label{IVb} 
Bulk properties of nuclear matter are very insightful for testing theories as well as providing a connection with bulk properties of nuclei.
On the other hand, momentum- and density-dependent single-particle potentials (SPP) in nuclear matter provide complementary, and more detailed information which is needed for HI transport simulations.

Together with the SPP in NM, one can construct the momentum and density dependent SPP in isospin-asymmetric matter -- and thus the symmetry potential -- to be used, for instance, in Boltzmann-Uehling-Uhlenbeck (BUU) calculations of collective nuclear dynamics. A number of HI collision observables have been found to be sensitive to the symmetry potential, such as the neutron/proton ratio of pre-equilibrium nucleon emission, neutron-proton differential flow, and the proton elliptic flow at high transverse momenta. 

\begin{figure*}[!t] 
\centering
\hspace*{-1cm}
\includegraphics[width=7.0cm]{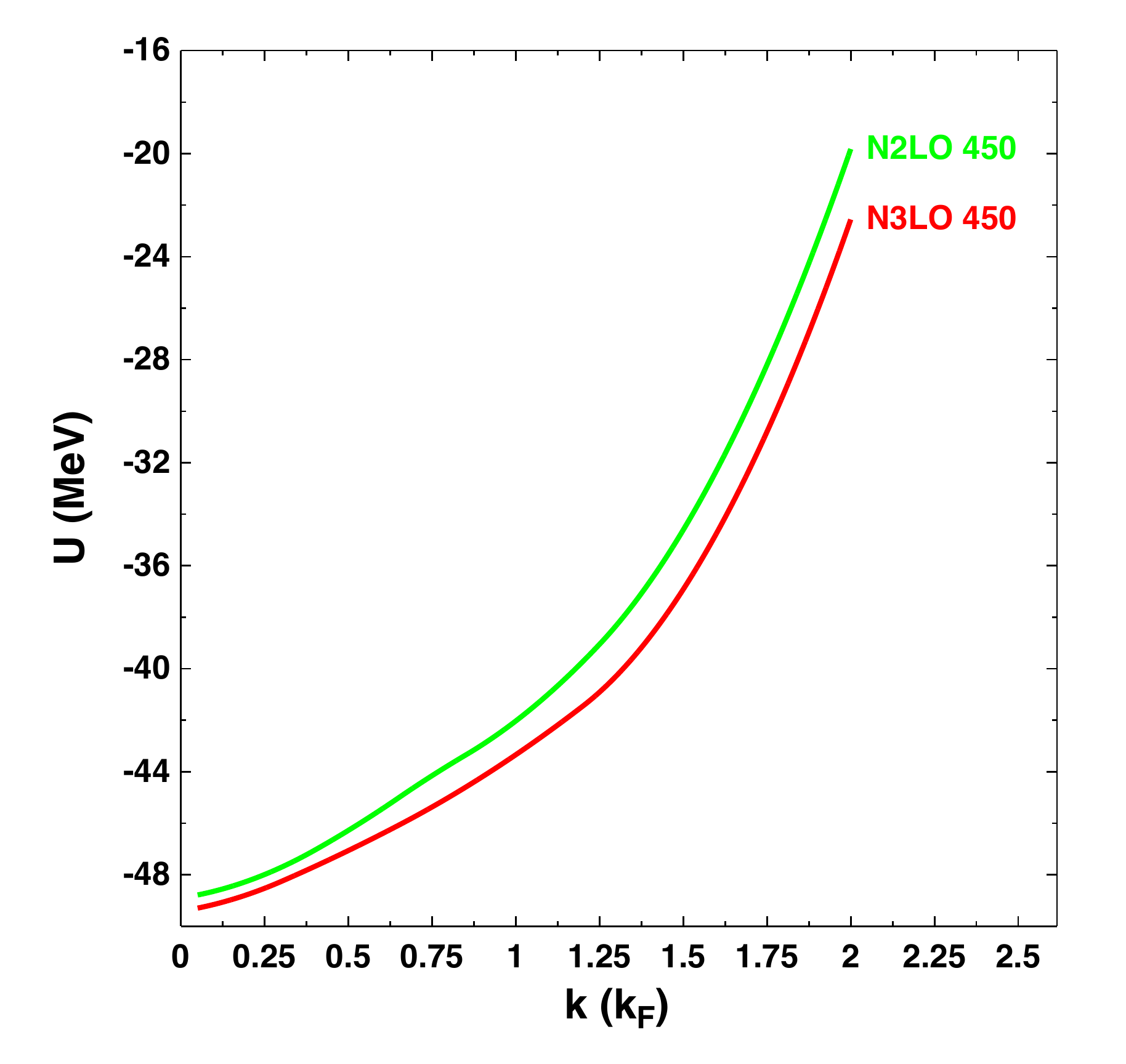}\hspace{0.01in} 
\includegraphics[width=7.0cm]{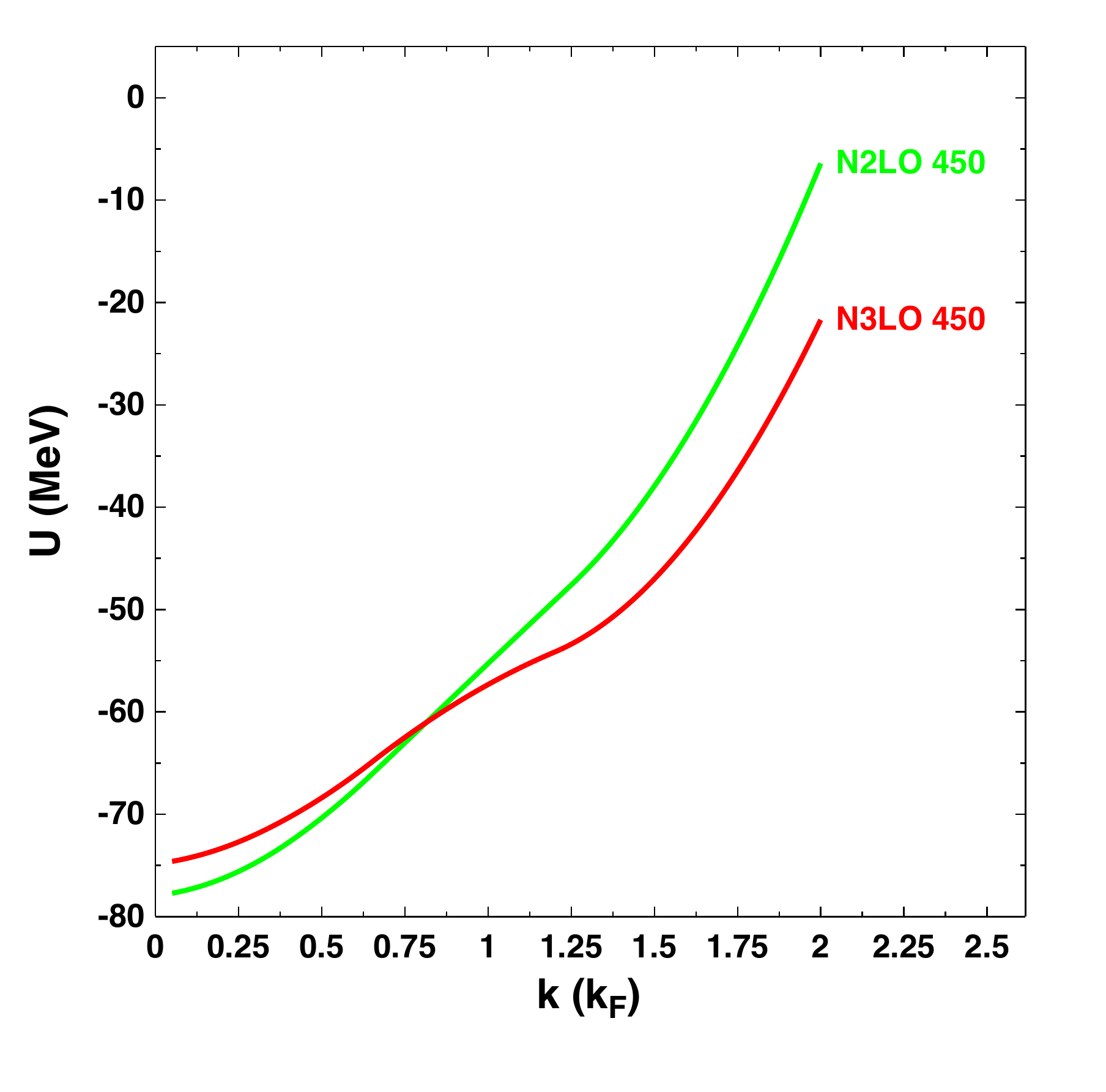}\hspace{0.01in} 
\vspace*{0.05cm}
\caption{Predictions for the SPP at N$^2$LO and N$^3$LO. The cutoff is fixed at 450 MeV. For the left (right) frame, the Fermi momentum is equal to $k_F$ = 1.0 (1.333) fm$^{-1}$.
}
\label{u_ord}
\end{figure*}

Next, we will take a look at the underlying Brueckner SPP, derived self-consistently with the $G$-matrix and, thus, the EoS, to learn about its momentum dependence and how that changes with density and chiral order.
First, for two selected densities (saturation density and about 2/3 of it, corresponding approximately to $k_F$ = 1.0 fm$^{-3}$), we show the single-particle potential at third and fourth order, Fig.~\ref{u_ord}. 

\begin{figure*}[!t] 
\centering
\hspace*{-1cm}
\includegraphics[width=7.0cm]{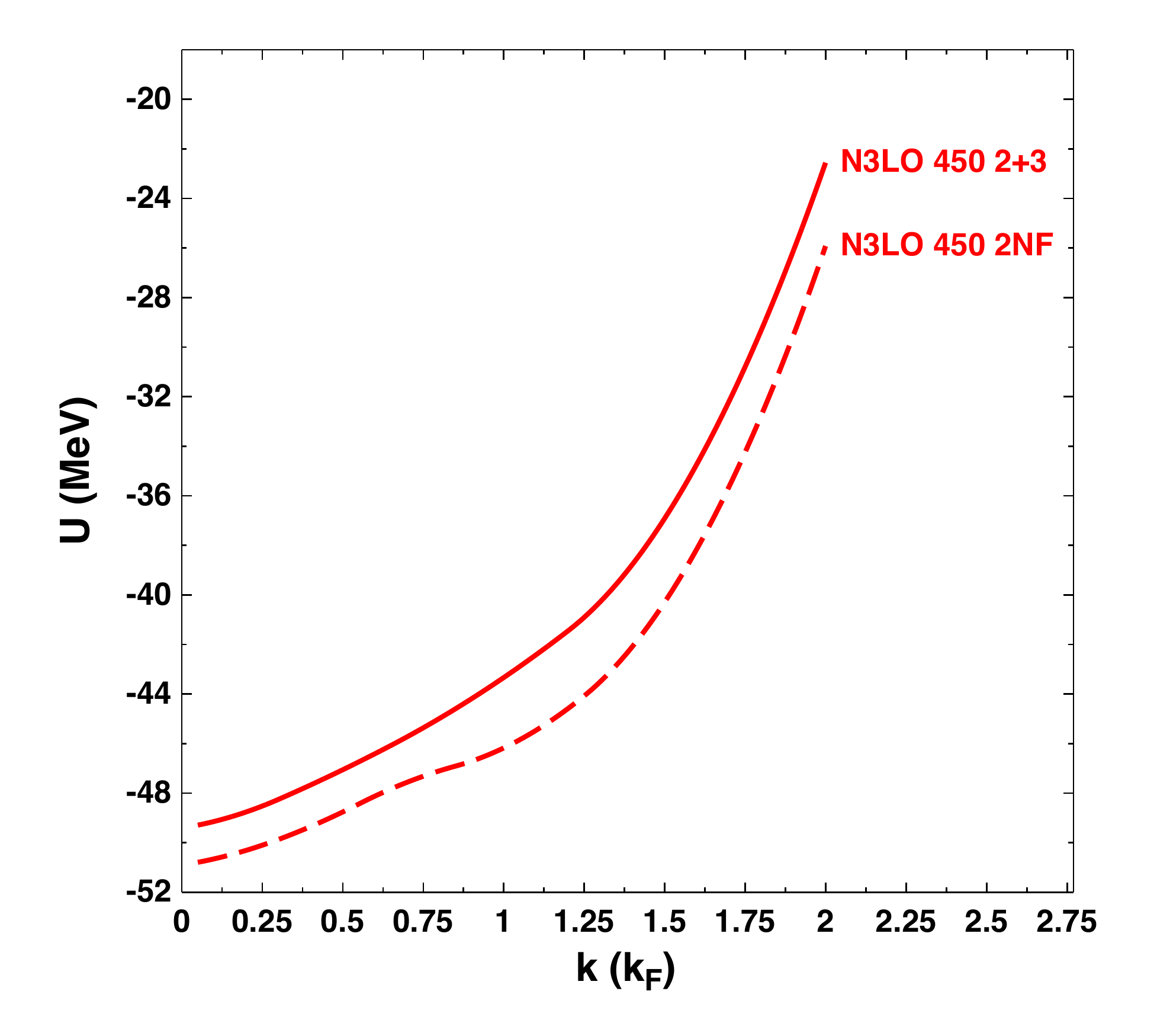}\hspace{0.01in} 
\includegraphics[width=7.0cm]{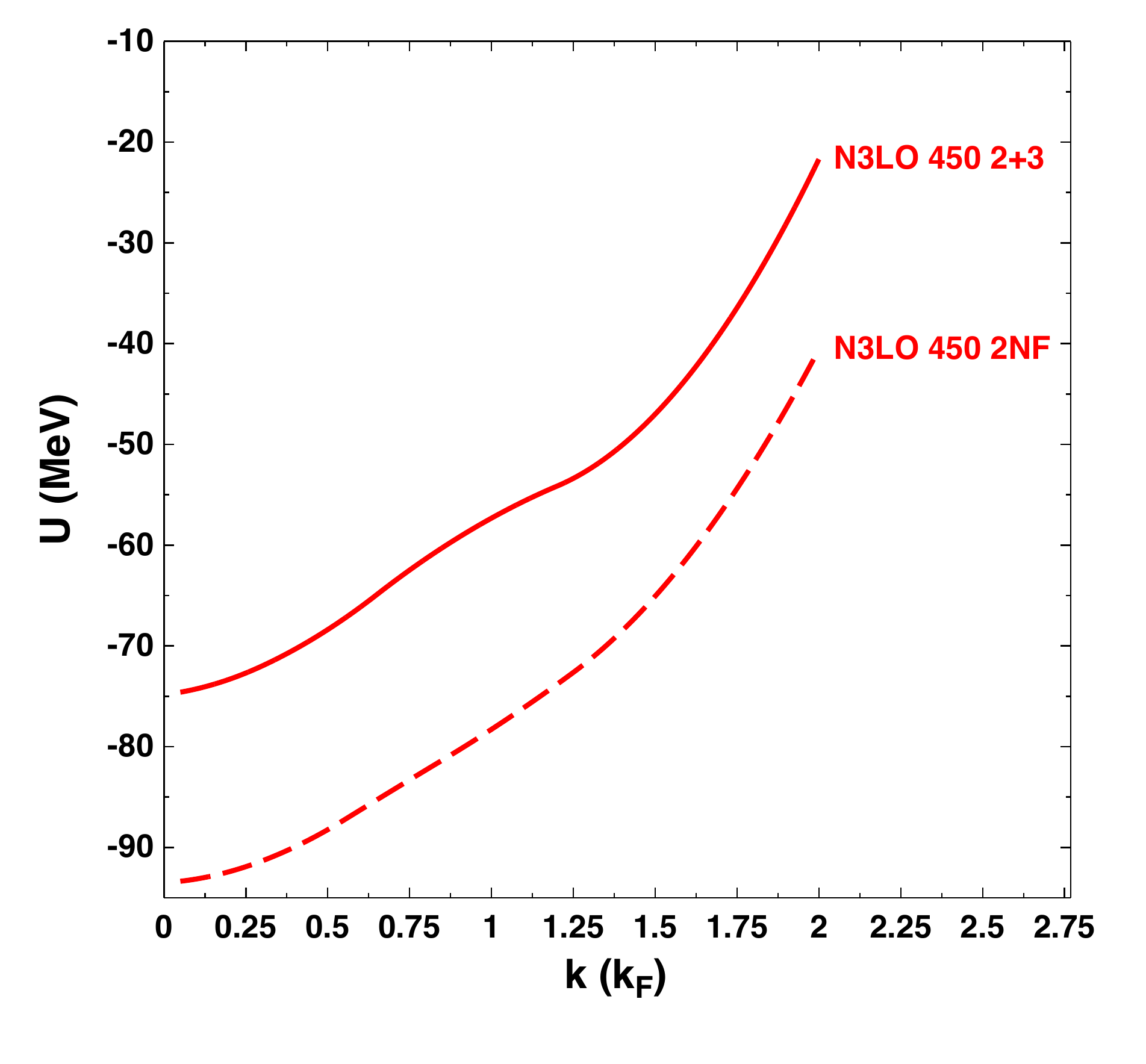}\hspace{0.01in} 
\vspace*{0.05cm}
\caption{Impact of including the 3NF up to N$^3$LO (solid curves) at two different densities. Left: $k_F$ = 1.0 fm$^{-1}$; right: $k_F$ = 1.333 fm$^{-1}$. The cutoff is fixed at 450 MeV.
}
\label{u_ord_23}
\end{figure*}   

Single-particle potentials derived from chiral interactions are generally deep and grow monotonically from the bottom of the Fermi sea.
  The impact of moving to fourth order is much larger at the higher density.

The impact of including the complete 3NF at N$^3$LO is demonstrated in Fig.~\ref{u_ord_23}. The effect is to decrease the depth of the potential, and is strongly density dependent.

Analyses of HI collision measurements are used to extract empirical constraints for the EoS. For instance, the elliptic flow in midperipheral to peripheral collisions was found to be particularly sensitive to the momentum dependence of the nucleon mean field~\cite{Dan20}.
 We suggest that extraction of reliable constraints to the EoS and/or the symmetry energy through analyses of reaction observables should be guided by state-of-the-art theories of nuclear forces.

\subsection{Short-range correlations} 
\label{IVb} 

Correlations in nuclear matter and nuclei carry important information about the 
underlying nuclear forces and their behavior in the medium. 
Since the early Brueckner nuclear matter calculations \cite{HT70}, it has been customary to associate the 
correlated two-body wave functions to the strength of the nucleon-nucleon $NN$ potential in specific channels.
 For instance, the $^3S_1-^3D_1$ channel will reveal tensor correlations, which 
is of particular interest,
 since the model dependence among predictions from different $NN$ potentials -- those which cannot be constrained by $NN$ data --  resides mostly    
in the strength of their respective (off-shell) tensor force. The most popular example is the deuteron D-state probability.

Here, we wish to look at some well-established concepts through a contemporary lens. First, a brief review of useful definitions.

In terms of relative and center-of-mass momenta, the Bethe-Goldstone equation can be written as 
\begin{displaymath}
G({\bf k}_0, {\bf k},{\bf P}^{c.m.}, E_0, k_F) = V({\bf k}_0, {\bf k}) +
\end{displaymath}
\begin{equation}
\int d^3{\bf k}^{'} V({\bf k}_0, {\bf k}^{'})        
\frac{Q(k_F,{\bf k}^{'}, {\bf P}^{c.m.})}{E-E_0}           
G({\bf k}^{'}, {\bf k},{\bf P}^{c.m.}, E_0, k_F) \; ,                                                                         
\label{BG} 
\end{equation}
where $V$ is the $NN$ potential, $Q$ is the Pauli operator, $E = E({\bf k}^{'}, {\bf P}^{c.m.})$, and 
$E_0 = E({\bf k_0}, {\bf P}^{c.m.})$, with the function $E$ the total energy of the two-nucleon pair. 

The second term of Eq.~(\ref{BG}) builds
SRC into the wave function through the infinite ladder sum. In operator notation, the correlated ($\psi$) and the uncorrelated ($\phi$) 
wave functions are related through 
\begin{equation}
G\phi = V \psi  \; , 
\label{wf} 
\end{equation}
from which it follows that 
\begin{equation}
\psi - \phi = V \frac{Q}{E-E_0}G\phi \; . 
\label{psi} 
\end{equation}
Equation~(\ref{psi}) defines the 
 difference between the correlated and the uncorrelated wave functions, 
$f=\psi - \phi$, referred to as the defect function. The defect function has the attribute of being different from zero over the (finite) range where SRC correlations are effective.   
          
It is convenient to consider the momentum-dependent Bessel transform of the defect function for each angular 
momentum state [and average center-of-mass momentum $P_{avg}^{c.m.}(k_0,k_F)$]:           
\begin{equation}
f_{LL'}^{JST}(k,k_0,k_F) = \frac{k \; \bar{Q}(k_F,k,P_{avg}^{c.m.}) G_{LL'}^{JST}(P_{avg}^{c.m.},k,k_0)}{E_0-E} \; , 
\label{fl} 
\end{equation}
where the angle-averaged Pauli operator has been employed. 
The magnitude squared of $f_{LL'}^{JST}(k,k_0,k_F)$ is the probability of exciting two nucleons with relative momentum $k_0$ and relative orbital
angular momentum $L$ to a state with 
relative momentum $k$ and relative orbital
angular momentum $L'$. (Following an earlier work~\cite{FS14}, we take 
the initial momentum equal to 0.55$k_F$.)
These components of the correlated wave function are the basis for the definition of the          
 ``wound integral", which,
 for each partial wave at some density $\rho$, is given by
\begin{equation}
\kappa_{LL'}^{JST}(k_0,k_F) = \rho \int_0^{\infty} |f_{LL'}^{JST}(k,k_0,k_F)|^2 dk \; .         
\label{kappa} 
\end{equation}
Thus, $f$ and $\kappa$ provide a clear measure of the strength of correlations present in each channel. The wound integral was first introduced by Brandow~\cite{Bra66} in the context of the Brueckner-Bethe-Goldstone theory of nuclear matter.

In Table~\ref{tab2}, we present the contributions to the integral, Eq.~(\ref{kappa}), from selected states or groups of states for different choices of the interaction and three densities. For all densities and models, it is apparent that SRC in nuclear matter are mainly due to coupled $S$-waves. At both the third and the fourth orders, the impact of 3NFs is largest in $^3S_1-^3D_1$ -- more so at the fourth order -- indicating additional tensor force from the 3NF. With regard to density dependence, several mechanisms play competing roles in the density dependence of $\kappa$, such as weaker Pauli blocking at lower density, enhanced impact of the repulsive core with increasing density (for partial waves dominated by the central force), increased strength of the tensor force from the 3NFs. Overall, looking at the values of $\kappa$ from all partial waves, we conclude that SRC generally decrease as density increases for the cases with only 2NFs, whereas the opposite is true in the presence of 3NFs -- possibly the result of competing effects from the 3NF (enhancing correlations) and Pauli blocking.

\begin{figure*}[!t] 
\centering
\hspace*{-1cm}
\includegraphics[width=8.5cm]{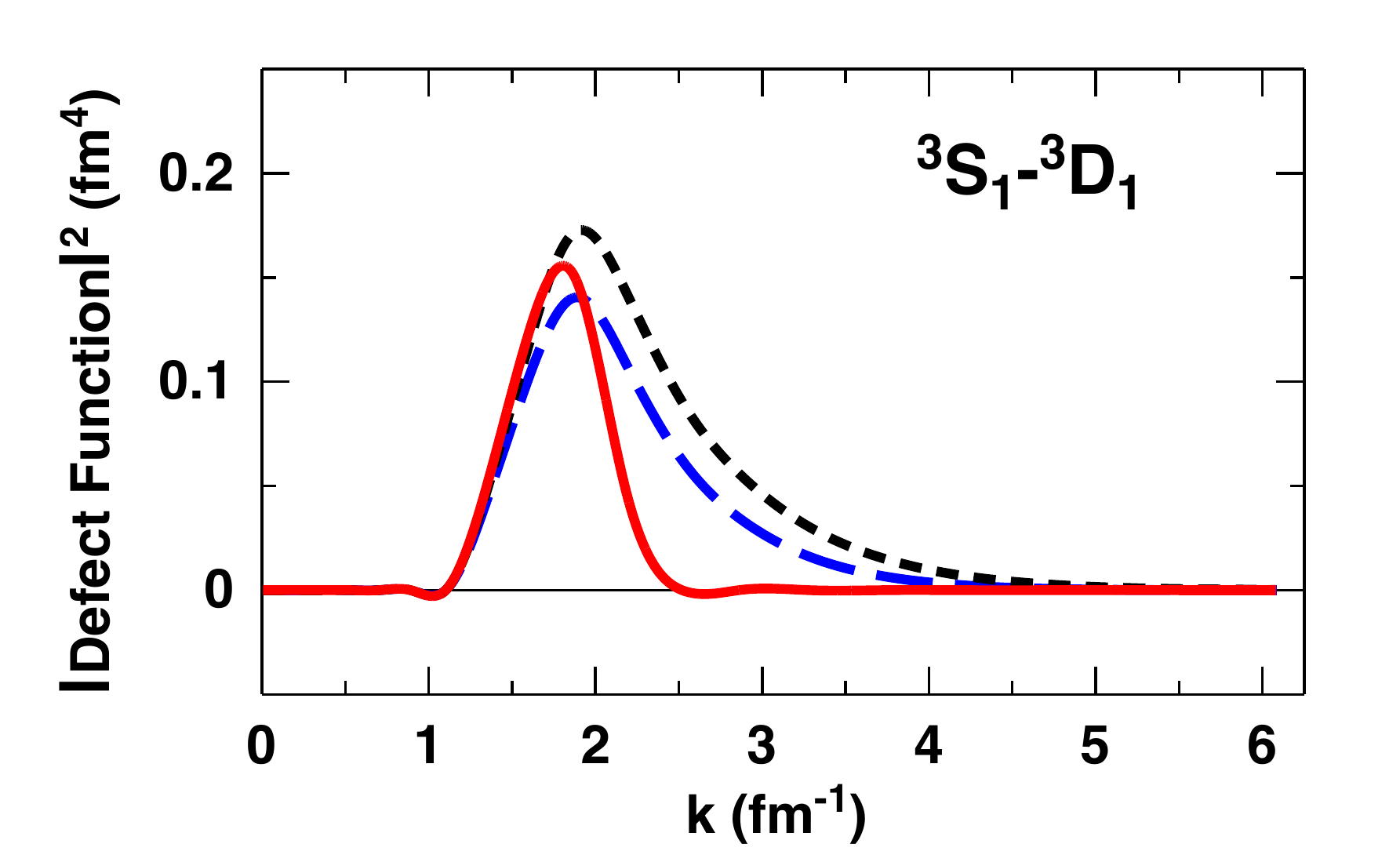}\hspace{0.01in} 
\vspace*{0.05cm}
\caption{ Magnitude squared of the defect function in the $^3S_1-^3D_1$ channel as a function of momentum. Solid red: N$^3$LO(450); dashed blue: CD-Bonn; dotted black: AV18. All curves are obtained with 2NFs only.
}
\label{def}
\end{figure*}

In Table~\ref{tab3}, we show the values of $\kappa$ (from all partial waves) obtained with three very different 2NFs: a state-of-the-art chiral potential, a high-precision momentum-space potential from the 90's~\cite{cdbonn}, and the local  AV18~\cite{WSS95}. In Fig.~\ref{def},  the probability amplitudes -- magnitude squared of Eq.(\ref{ff})  for the $J=1$ coupled states-- are displayed for the three cases considered in Table~\ref{tab3}. The impact of the cutoff  in chiral EFT is apparent, with the local AV18 extending the farthest, and both AV18 and CD-Bonn extending much farther than N$^3$LO.

\begin{table*} [t]               
\centering \caption                                                    
{Contributions to the wound integral, Eq.~\ref{kappa}, from $J=0$ and $J=1$ states with different interactions and changing densities. The last column shows the contribution from all partial waves.
} 
\vspace{5mm}
\begin{tabular*}{\textwidth}{@{\extracolsep{\fill}}cccccccc}
\hline
$ k_F$ (fm$^{-1}$) & Model &  $^1S_0$ & Total from J=0&  $^3S_1-^3S_1$ &  $^3S_1-^3D_1$ &  Total from J=1 & All partial waves \\
\hline     
 1.1 &  N$^2$LO & 0.0081& 0.0086 & 0.027 & 0.047 & 0.087 & 0.093 \\
       &  N$^2$LO+3NF & 0.0028 & 0.0033 & 0.015 & 0.079 & 0.1064 & 0.1141 \\
      &  N$^3$LO &  0.011 & 0.011 & 0.038 & 0.062 & 0.1146 & 0.1203 \\
       &  N$^3$LO+3NF & 0.0088 & 0.0092 & 0.0351 & 0.095 & 0.1479 & 0.1555 \\
\hline
 1.3 &  N$^2$LO & 0.0033 & 0.0039 & 0.011& 0.033 & 0.053 &0.059 \\
       &  N$^2$LO+3NF & 0.0054& 0.0059 & 0.0040& 0.082 & 0.109 & 0.1185 \\
      &  N$^3$LO & 0.0085 & 0.0088 & 0.019 & 0.046 &0.079 & 0.085 \\
       &  N$^3$LO+3NF & 0.016 & 0.016 & 0.015 & 0.096 & 0.141 & 0.151 \\
\hline
 1.4 &  N$^2$LO & 0.0023 & 0.0031 & 0.0070 &0.027 & 0.042 & 0.048 \\
       &  N$^2$LO+3NF & 0.0095 &0.010 & 0.0047 & 0.084 & 0.1194 & 0.1303 \\
      &  N$^3$LO & 0.0090 & 0.0093 & 0.014 & 0.038 & 0.067 & 0.073 \\ 
       &  N$^3$LO+3NF & 0.022 & 0.023 & 0.0094 & 0.097 & 0.1484 & 0.1599 \\
\hline
\end{tabular*}
\label{tab2} 
\end{table*}

\begin{table*} [t]               
\centering \caption                                                    
{The wound integral, Eq.~\ref{kappa}, from three different interactions around saturation density. In each case, only 2NFs are included.
} 
\vspace{5mm}
\begin{tabular*}{\textwidth}{@{\extracolsep{\fill}}ccc}
\hline
$ k_F$ (fm$^{-1}$) & Model  &  Contribution to $\kappa$ from all partial waves \\
\hline
 1.3  &  N$^3$LO(450) &  0.085 \\
       &  CD-Bonn & 0.114 \\
      & AV18 & 0.157 \\
\hline
\end{tabular*}
\label{tab3} 
\end{table*}

These quantities, which can be dramatically different from model to model -- as has been known for decades --  are not observable. The SRC probabilities and high-momentum distributions in nuclei, which have been and are being extracted from hard electron scattering experiments~\cite{CLAS,CLAS2,CLAS3,Pia+,Eg+06,Shneor,Subedi,Baghda,Pia13,Korover,Hen+17,CT+,Atkwim19} are equally non-observable, although high-momentum information can be extracted from data in a scale and scheme dependent way~\cite{TBF21}.
The recent comprehensive analysis from Ref.~\cite{TBF21} describes the situation very clearly: the SRC knock-out experiments do have merit, but their value ``{\it  ...is not new insight into the interaction, but to demonstrate that short-range physics can be isolated and to a certain extent controlled.}" Results of these experiments cannot be used to to select the ``best" off-shell behavior, a concept that can be proven to be fundamentally impossible~\cite{HF98,FH02,TBF_arx}. For instance, the momentum distribution of AV18 extends past 4 fm$^{-1}$, meaning that  strong SRC are built into the wave function. On the other hand of the spectrum are SRG-evolved interactions, with no high-momentum components. If predictions with a particular potential are closest to the knock-out measurements, in no way that implies that the ``measured" off-shell behavior has selected that particular interaction -- it means that the latter is more suitable for the assumptions made in the data analyses, for instance, impulse approximation.
Ultimately, predictions from observables must agree for any realistic model, whether SRC are built into the wave function or in the operators~\cite{TBF21}.

\section{Summary, conclusions, and future plans}                                                                  
\label{Concl} 

We calculated the EoS of SNM from leading to fourth order. At N$^3$LO, we include all subleading 3NFs.
 An EoS with good saturation properties (density, energy, and curvature) can be obtained from chiral EFT and a softer cutoff (smaller than 500 MeV). 

We have also shown a representative sample of SPP results, which we obtain self-consistently from the $G$-matrix. We find the effect of 3NFs on the SPP to be large at normal densities. 
Microscopically calculated SPP provide useful information  to guide parametrizations of the nucleon potential for use in transport simulations. 

We then moved to a discussion of SRC in nuclear matter, as seen through the momentum-space defect function and the integral of its magnitude squared. Central and tensor correlations are seen mostly in uncoupled and coupled $S$-waves. We took the opportunity to comment on the model dependence and the non-observable nature of SRC.

Having the EoS for SNM and NM~\cite{SM21} consistently at N$^3$LO, we are in the position to revisit neutron skins and neutron stars.
Our work in progress includes another form of correlations, namely pairing in nuclear and neutron matter. Pairing is a two-body correlation near the Fermi surface -- hence, it has features common to any quantum system of fermions.  The appearance of superfluidity in neutron stars suppresses standard neutrino cooling processes, and thus pairing effects can have a remarkable role on the evolution of a neutron star.

\section*{Acknowledgments}
Support by the U.S.\ Department of Energy, Office of Science, Office of Basic Energy Sciences, under Award Number DE-FG02-03ER41270 is acknowledged.

\end{document}